\documentclass[a4paper,11pt]{article}
\pdfoutput=1

\usepackage{jinstpub}

\usepackage{listings}
\PassOptionsToPackage{hyphens}{url}
\appto\UrlBreaks{\do\/}
\appto\UrlBigBreaks{\do\/}

\usepackage{lineno}
\usepackage{cprotect}
\newcommand{\wzonline}[1]{}

\newcommand{\wzurl}[1]{\href{#1}{\url{#1}}}

\usepackage[frozencache=true,cachedir=minted-cache]{minted}

\title{\boldmath Easy and structured approach for software and firmware co-simulation for bus centric designs}

\author{Michał Kruszewski}

\affiliation{Institute of Electronic Systems, Warsaw University of Technology,\\
    Nowowiejska 15/19, 00-665 Warszawa, Poland}

\emailAdd{michal.kruszewski@pw.edu.pl}

\abstract{
Although software and firmware co-simulation is gaining popularity, it is still not widely used in the FPGA designs.
This work presents easy and structured approach for software and firmware co-simulation for bus centric designs.
The proposed approach is very modular and software language agnostic.
The only requirement is that the firmware design is accessible via some kind of system bus.
The concept has been used for testing DAQ system being developed for high energy physics experiment.
}


\proceeding{TWEPP 2021\\
 20 -- 24.09.2021}

\begin{document}
\maketitle
\flushbottom

\section{Introduction}

Software and firmware co-simulation can save a lot of time and money, as it leads to a lower number of HDL project builds and reduces the number of test iterations with the real hardware.
Despite its apparent advantages the co-simulation is still relatively rare to see in FPGA designs.
There are at least four reasons for such a situation.
The first one is that setting up a co-simulation framework requires knowledge of multiple computing areas.
The second one is that it might be time-consuming.
The third one is that ready-to-use frameworks sometimes do not support some of the desired HDL features, for example handling compound types such as records or arrays.
The fourth one is that ready-to-use frameworks, such as cocotb \cite{noauthor_cocotbcocotb_2021}, are strictly coupled with a single programming language.
This work presents the modular approach that tries to be in line with the Unix philosophy.

\section{Concept}

\begin{figure}[!t]
	\centering
	\includegraphics[scale=0.60]{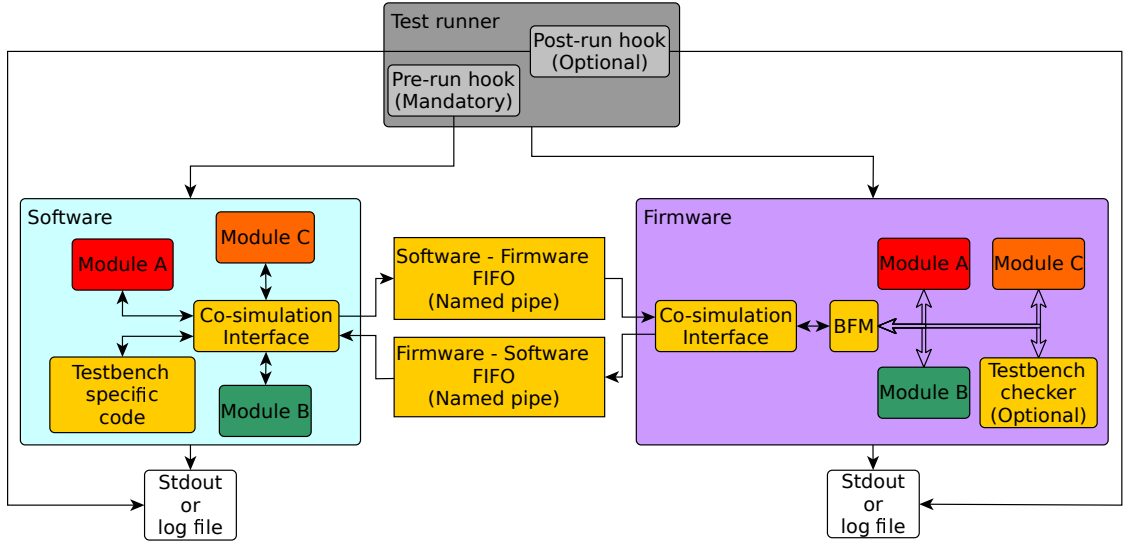}
	\caption{Scheme of the co-simulation framework concept.}%
	\label{fig:scheme}
\end{figure}

The co-simulation framework proposed in this work consists of the following mandatory elements:
\begin{enumerate}
	\item software co-simulation interface,
	\item HDL co-simulation interface,
	\item HDL BFM (Bus Functional Model),
	\item IPC (Inter-Process Communication) mechanism,
	\item test runner.
\end{enumerate}

Figure \ref{fig:scheme} shows scheme of the co-simulation framework concept.
The framework blocks are loosely coupled, and each of them can be easily replaced.
The whole test bench additionally consists of the project’s software and firmware code.
The co-simulation interfaces are relatively short and straightforward, and once written, they can be reused for different tests within the project.
If different software languages are used for the prototype and target implementation phases, it is also easy to write a co-simulation interface for the new language and reuse the co-simulation framework from the prototyping phase.
The BFM can be custom or taken from a library such as OSVVM \cite{noauthor_osvvm_2021} or UVVM \cite{uvvm_uvvm_2021, tallaksen_uvvm_2018}.
The idea is based on the assumption that all communication is done via the bus.
Not only the regular data is transferred via the bus, but also the test bench specific data.
Such an approach is immune to the lack of support for compound types.
What is more, the bus infrastructure is tested by the way.

The proposed approach is not free of drawbacks.
The first one is that the firmware design must have some kind of system bus.
This should not be a problem as almost all complex FPGA designs have some kind of bus, Wishbone \cite{opencores_wishbone_nodate} and AXI \cite{arm_amba_nodate} being probably the most popular.
The second one is that precise timing checking between signals is hard to achieve solely within the test bench software.
It requires a firmware checker accessible via the bus.
Another approach is using PSL (Property Specification Language) or SVA (SystemVerilog Assertions).

\section{Implementation}

The concept has been successfully implemented.
A simplified example showing a co-simulation for an adder is available on \cite{mkru_fusesoc_2021}.
The FuseSoc \cite{kindgren_fusesoc_2021, kindgren_invited_2019} and fsva \cite{mkru_fsva_2021} tools have been used as the test runner.
The AGWB \cite{zabolotny_agwb_2021, zabolotny_automatic_2019} tool is used for the registers generation.
Wishbone has been chosen as the system bus.
The firmware bus infrastructure comes from the General-Cores \cite{ohwr_gc_nodate} library.
The Wishbone BFM comes from the UVVM library.
Although all listed components are necessary, it is worth noting that the whole concept is agnostic to the chosen components.
They all depend on the project and personal preferences.

\begin{figure}[!t]
	\centering
	\includegraphics[scale=0.40]{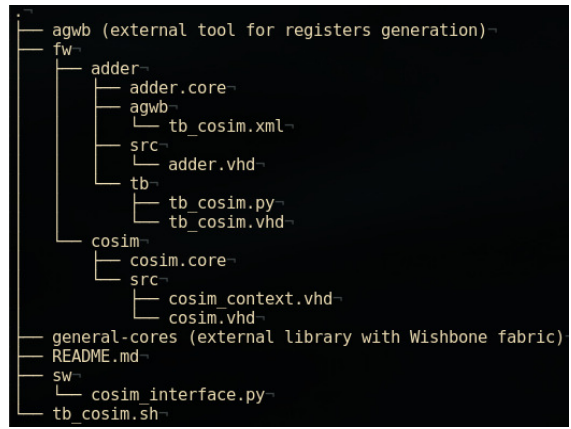}
	\caption{Directory structure of the exmaple co-simulation project.}%
	\label{fig:tree}
\end{figure}

Figure \ref{fig:tree} shows directory structure of the example.
At first it may seem that setting up single co-simulation requires relatively a lot of files.
However, most of these files are reused between co-simulations and are short.
For example, files: \emph{tb\_cosim.sh}, \emph{sw/cosim\_interface.py} and all files in the \emph{fw/cosim} directory are reused in case of multiple co-simulations in the same project.
The only two files stricly related with the particular module co-simulation are \emph{tb\_cosim.vhd} and \emph{tb\_cosim.py}.

\begin{listing}[h]
	\begin{minted}[fontsize=\scriptsize]{yaml}
targets:
  tb_cosim:
    default_tool: ghdl
    toplevel: tb_cosim
    generate:
      - agwb_regs
    filesets:
      - agwb_dep
      - src
      - tb_cosim
    hooks:
      pre_run: [tb_cosim]
    parameters:
      - G_SW_FW_FIFO_PATH=/tmp/fusesoc_cosim_example/adder_python_vhdl
      - G_FW_SW_FIFO_PATH=/tmp/fusesoc_cosim_example/adder_vhdl_python

parameters:
  G_SW_FW_FIFO_PATH:
    datatype: str
    paramtype: generic
  G_FW_SW_FIFO_PATH:
    datatype: str
    paramtype: generic

scripts:
  tb_cosim:
    cmd:
      - ../../../tb_cosim.sh
      - adder
      - /tmp/fusesoc_cosim_example/adder_python_vhdl
      - /tmp/fusesoc_cosim_example/adder_vhdl_python
	\end{minted}
	\caption{adder.core file snippet showing passing paths for named pipes.}
	\label{lst:adder_core}
\end{listing}

Listing \ref{lst:adder_core} presents the snippet from \emph{adder.core} file, showing how paths for named pipes are passed both to the firmware and software sides.
The paths could be hardcoded in the \emph{tb\_cosim.vhd} and \emph{tb\_cosim.py} files, however such approach increases the maintenance burden and enforces keeping the same information in multiple files.
Keeping all the paths related information in the single \emph{.core} file also makes any further editing easier.

\subsection{Co-simulation interface}

The co-simulation interface is a custom protocol for controlling the BFM and simulation progress.
Its features and structure depend on the particular project requirements.
The bare minimal interface for two-side communication must support write and read bus transfers, as well as a command for advancing a simulation for a given amount of time.
More complex interfaces might also support block read, block write and stream transfers.
They can also model access times or count the number of transactions.

\section{Example run and output}

\begin{figure}[!t]
	\centering
	\includegraphics[scale=0.40]{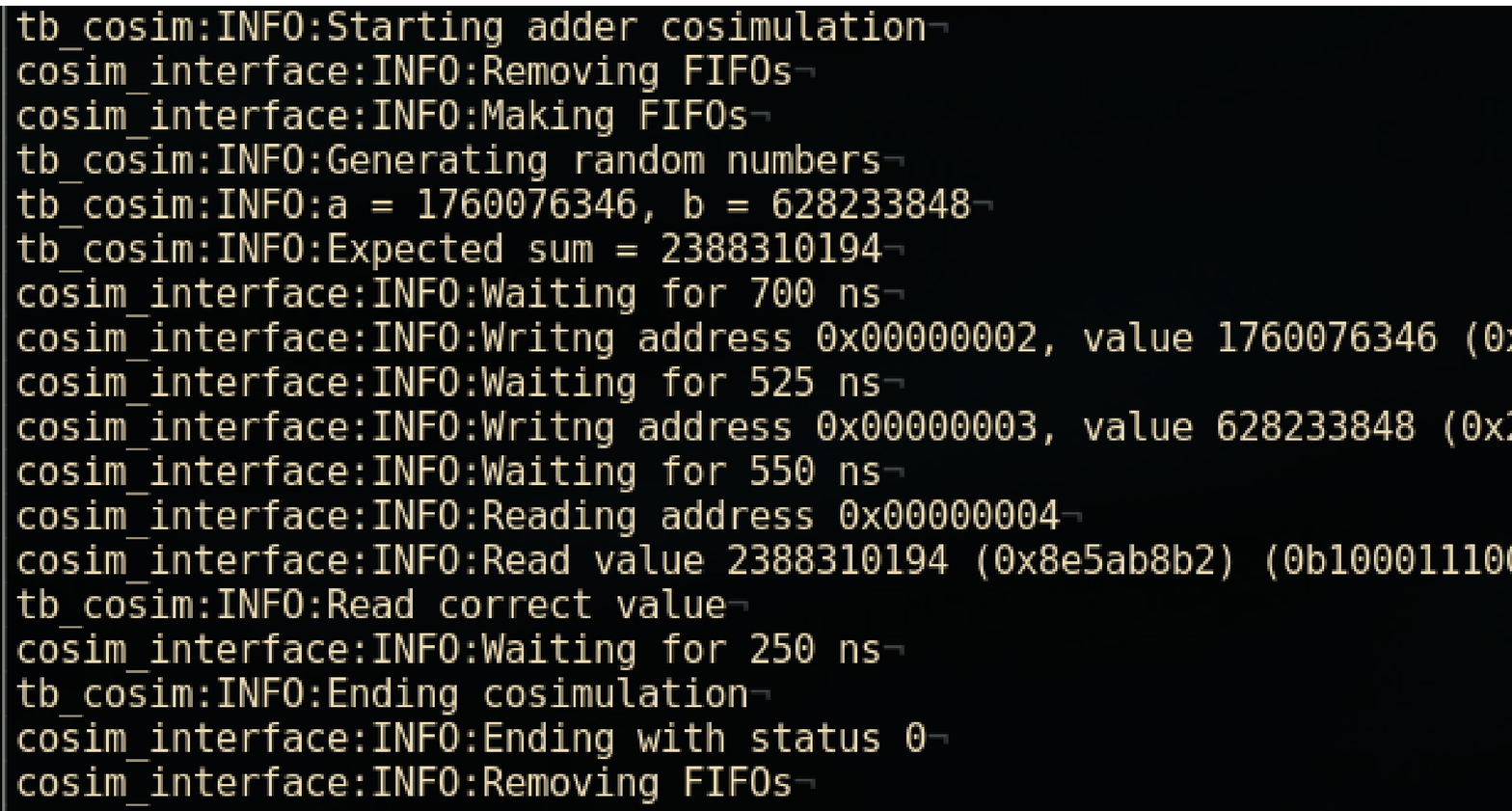}
	\caption{Software side log for the example co-simulation.}%
	\label{fig:sw_log}
\end{figure}

\begin{figure}[!t]
	\centering
	\includegraphics[scale=0.37]{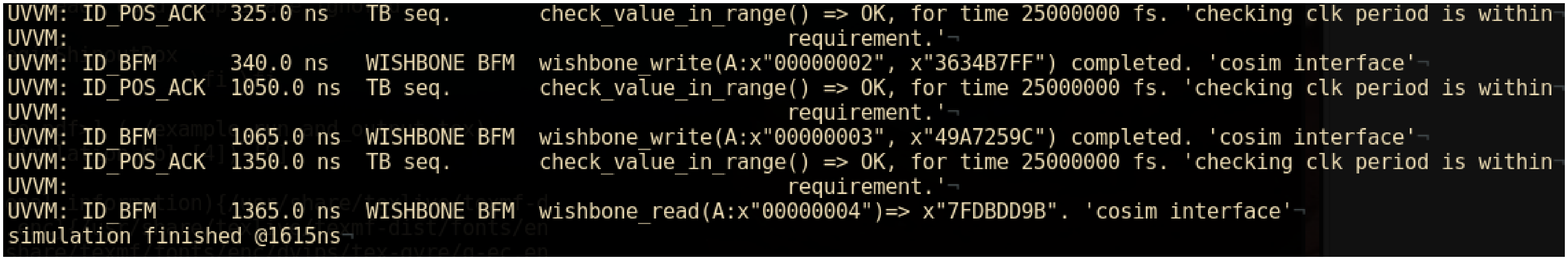}
	\caption{Firmware side log for the example co-simulation.}%
	\label{fig:fw_log}
\end{figure}

To run the co-simulation in the example project one needs to simply execute \lstinline{fsva ::adder tb_cosim} (assuming the dependencies listed in the \emph{README.md} file are already installed).
By default, only the standard output and standard error from the firmware side are attached to the terminal.
This is because only the firmware simulator is started directly by the FuseSoc.
The software is started by the pre-run hook, and its standard output and standard error need redirecting to a file.
This is done in the \emph{tb\_cosim.sh} file.
A need to redirect the output from the software side is not a big issue, as one can simply run \lstinline{tail -f /tmp/fusesoc_cosim_example/adder.log} to get a live, terminal like print experience.
Figure \ref{fig:sw_log} presents software side log and figure \ref{fig:fw_log} presents firmware side log.

\section{Real use case}

The proposed approach has been used for testing of DAQ (Data Acquisition) system for the CBM (Compressed Baryonic Matter) \cite{noauthor_compressed_2012} experiment that is being prepared at FAIR (Facility for Antiproton and Ion Research) in Darmstadt.

\acknowledgments
The work has been partially supported by the statutory funds of the Institute of Electronic Systems.

\bibliography{fusesoc_cosimulation}
\bibliographystyle{JHEP}

\end{document}